\def\ARXIV{}

\ifdefined\ARXIV
  \documentclass[sigconf,nonacm]{acmart}
\else
  \documentclass[sigconf,anonymous,review]{acmart}
\fi

\usepackage{amsmath}
\usepackage{booktabs}
\usepackage{array}
\usepackage{graphicx}
\usepackage{xspace}

\ifdefined\ARXIV
  \setcopyright{none}
\else
  \setcopyright{none}
  \copyrightyear{2027}
  \acmConference[WSDM '27]
    {The Twentieth ACM International Conference on Web Search and Data Mining}
    {February 15--19, 2027}
    {Hong Kong, China}
  \acmYear{2027}
  \acmISBN{}
  \acmDOI{}
\fi

\newcommand{\R}{\mathbb{R}}

\newcommand{\mean}{\operatorname{mean}}
\newcommand{\ndcg}{\ensuremath{\mathrm{NDCG@10}}\xspace}

\title{What Score Distillation Transfers to Compact Dense Retrievers}

\ifdefined\ARXIV
  \author{Kirill Dubovikov}
  \affiliation{
    \institution{Mohamed bin Zayed University of Artificial Intelligence}
    \city{Abu Dhabi}
    \country{United Arab Emirates}
  }
  \email{kirill.dubovikov@mbzuai.ac.ae}

  \author{Martin Tak\'a\v{c}}
  \affiliation{
    \institution{Mohamed bin Zayed University of Artificial Intelligence}
    \city{Abu Dhabi}
    \country{United Arab Emirates}
  }
  \email{martin.takac@mbzuai.ac.ae}

  \author{Salem Lahlou}
  \affiliation{
    \institution{Mohamed bin Zayed University of Artificial Intelligence}
    \city{Abu Dhabi}
    \country{United Arab Emirates}
  }
  \email{salem.lahlou@mbzuai.ac.ae}
\else
  \author{Anonymous Authors}
  \affiliation{
    \institution{Anonymous Institution}
    \country{}
  }
  \email{}
\fi

\begin{document}

\begin{abstract}
Compact retrievers are often specialized by distilling a large embedding
model's similarity scores on hard negatives mined by the teacher model. We
argue that this common setup calls for a different loss than the
scale-dependent ones usually used, such as MarginMSE and PairMSE. When the teacher discards candidates scored too close to the
labeled positive, the negatives it keeps get nearly identical scores. These
losses then mostly ask the student to reproduce this tiny score range rather
than the teacher's ranking. A per-row Pearson correlation loss, invariant to
the shift and scale of the scores in each row (a query and its 32 candidates),
avoids this failure, and removing
PairMSE's scale term alone recovers most of Pearson's lead. We propose
\emph{row scale dispersion}, computed from the untrained student before
training, to tell when this matters. With this loss, on
rows mined around a labeled document, the teacher's scores substitute for
relevance labels rather than add to them. Training on the binary label alone works as
well, with the same loss or, for one student, a contrastive loss: the
teacher's value lies in choosing the negatives. Without labels, the teacher's
scores on its own top-ranked documents recover most of the labeled gain,
largely through its top document as a pseudo-label. Experiments with two compact
students on three BEIR datasets support both conclusions, including, for
one student, comparisons with MarginMSE and a tuned listwise KL. We suggest a
simple practical rule: distill teacher-mined negatives
with a scale-free loss and, on such rows, use the teacher's scores where labels
are missing rather than on top of them.
\end{abstract}

\ifdefined\ARXIV\else

\ccsdesc[500]{Information systems~Retrieval models and ranking}
\ccsdesc[300]{Computing methodologies~Supervised learning}

\keywords{dense retrieval, embedding models, knowledge distillation, model compression}
\fi

\maketitle

\section{Introduction}

Large embedding models such as NV-Embed and Qwen3 Embedding perform high quality
retrieval \citep{lee2024nvembed,qwen2025embedding}, but every query must pass through a
multi-billion-parameter encoder at serving time, and every document at indexing
time; a compact student specialized to a deployment's known collections could
serve them far more efficiently.

Distilling a large embedding teacher's similarity scores into a smaller
embedding model is established practice
\citep{awasthy2025granite,zhang2024jasper}, while our work analyzes what
transfers from teacher to student. In our setup, the teacher scores a \emph{row} of 32 candidate documents per
query offline; the student learns these cosine scores with RD-Suite's PairMSE
loss \citep{qin2023rdsuite} or a per-row Pearson loss and is then served
alone. For Qwen3-Embedding-0.6B \citep{qwen2025embedding} and E5-large-v2
\citep{wang2022e5}, we vary how rows are built (target or other collections,
TF-IDF or teacher-mined candidates, with or without labels), the loss, and the
target (the teacher's scores, the binary label alone, or corrupted scores). We
evaluate on three target collections that supply the training rows (SciFact,
NFCorpus, FiQA) and five excluded collections never used in training. On TF-IDF rows, distillation specializes both students at
unchanged serving cost (0.018 \ndcg, a fifth of the gap to the teacher), and how the training rows are built determines what the
teacher's scores can teach. Our contributions are:
\begin{itemize}
  \item \emph{A loss for teacher-mined rows.} Mining leaves the teacher's
  scores nearly tied, and PairMSE asks the student to match their small
  spread, so the student gains little. Splitting PairMSE into scale and shape
  terms, we show that a per-row Pearson loss avoids this for both students
  and, for Qwen, also beats MarginMSE and a tuned listwise KL. Removing
  PairMSE's scale term alone recovers most of Pearson's lead.
  \item \emph{A diagnostic to choose the loss before training.} We propose
  \emph{row scale dispersion}: the spread over rows of the untrained
  student's score scale relative to the teacher's, from one scoring pass. It
  indicates when a scale-free loss is needed: it is high on mined rows, where
  Pearson wins, low on TF-IDF rows, where the losses nearly tie, and
  intermediate on rows sampled across the teacher's ranking, where the gap
  is small.
  \item \emph{What the teacher's scores are for.} On rows the teacher mines
  around the label, its scores substitute for relevance labels rather than add
  to them: a one-hot label target does as well. Without labels, the teacher's
  scores on its own top documents keep most of the gain, mostly through its
  top document.
\end{itemize}

\paragraph{Code availability.}
Source code, executable paper configurations, and aggregate result artifacts
are available at
\ifdefined\ARXIV
\url{https://github.com/kdubovikov/score-vector-distillation}.
\else
\url{https://anonymous.4open.science/r/score-vector-distillation-1245/}.
\fi

\section{Related Work}

\paragraph{Label-free distillation.}
Transfer sets may be unlabeled \citep{bucilua2006model,hinton2015distilling};
retrieval work replaces annotated pairs with reader attention
\citep{izacard2021reader}, question reconstruction \citep{sachan2023art},
teacher scores on generated queries
\citep{wang2022gpl,saadfalcon2023udapdr,tamber2025listwise}, or LLM data
\citep{bonifacio2022inpars,dai2023promptagator,lee2024gecko}.

\paragraph{What the student takes from the scores.}
Ranking distillation has often assumed that only the top of the teacher's list
matters \citep{tang2018ranking,reddi2021rankdistil}, while other work finds
signal in moderately relevant negatives \citep{tao2024adam} and across the full
score range: \citet{jang2026beyond} find that rows of the teacher's
top-ranked candidates often distill worse than rows sampled across its score
range, using scale-dependent losses (KL, MarginMSE).
\citet{qin2023rdsuite} ask where the useful signal in a teacher's score
vector lies; we locate it by corrupting or replacing the negatives' scores and by
removing the label from the rows.

\paragraph{Losses and teachers.}
Retrieval distillation uses positive-negative margins
\citep{hofstatter2020marginmse}, in-batch teacher supervision
\citep{hofstatter2021tasb}, joint reranker training \citep{ren2021rocketqav2},
pairwise preferences \citep{huang2024pairdistill}, curricula
\citep{zeng2022cldrd}, dual-encoder objectives \citep{menon2022defense}, and
late-interaction teachers \citep{macavaney2020tctcolbert}. Teacher-mined hard
negatives are standard, with positive-aware filtering against false negatives
\citep{moreira2024nvretriever}. Losses that match logits up to scale, by
Pearson correlation or standardization, exist in classification distillation
\citep{huang2022dist,sun2024logitstd}, and a Pearson loss was marginally better
than a Huber loss in text-to-video retrieval \citep{tian2023teachclip}; we
show when and why score scale matters for retrieval score distillation. Representation
alignment distills embeddings or hidden layers
\citep{kim2023embeddistill,vu2026mol}, also from black-box teachers without
judgments \citep{leaf2025teacheraligned} or through a projection
\citep{akram2026jina}; Search-Adaptor adapts a large model's embeddings
\citep{yoon2024searchadaptor}. Score distillation from embedding teachers
builds Granite Embedding \citep{awasthy2025granite} and Jasper and Stella
\citep{zhang2024jasper}. Closest to our setting, \citet{tamber2025listwise}
specialize BERT-base models to corpora including SciFact, NFCorpus, and FiQA,
where listwise KL from a cross-encoder on generated queries helps more
consistently than InfoNCE. We distill a black-box embedding teacher on real
target queries and analyze what transfers: which rows, which loss, whether
labels help, and which parts of the score vector matter. We also average two
teachers' scores per document (teacher fusion), which relates to multi-teacher
distillation
\citep{saadfalcon2023udapdr,lu2024mta4dpr} and to score fusion in metasearch
\citep{aslam2001models,montague2001normalization,dwork2001rank}.

\section{Score Distillation over Candidate Rows}
\label{sec:method}
For each query $q$, a candidate row contains $k=32$ documents
$d_0,\ldots,d_{k-1}$: a labeled positive $d_0$ and 31 negatives (rows
without labels are defined in Section~\ref{sec:setup}). The student assigns
scores $s_i=f_\theta(q,d_i)$ and the teacher assigns $t_i$; both are cosine
similarities of L2-normalized embeddings without further scaling, and teacher
and student may use different embedding spaces. Fusing two teachers
averages their raw cosine scores, without rescaling, to form $t$.

\paragraph{Row-centered loss.}
The loss is RD-Suite's PairMSE \citep{qin2023rdsuite}, which matches every
pairwise score difference in the row; we compute it in a row-centered form.
Shifting every candidate's score by the same amount does not change the
ranking, so we remove the mean student-teacher difference. With residual
$r=s-t \in \R^k$ and all-ones vector $\mathbf{1}_k$,
\begin{equation}
\mathcal{L}_{\mathrm{center}}(s,t)
= \frac{1}{k}\left\|r - \mean(r)\mathbf{1}_k\right\|_2^2 .
\label{eq:center-loss}
\end{equation}
Since \((s_i-s_j)-(t_i-t_j)=r_i-r_j\) and
\(\sum_{i,j}(r_i-r_j)^2=2k\sum_i(r_i-\mean(r))^2\), PairMSE averaged over all
$k^2$ ordered pairs equals twice the centered loss, which therefore needs no
explicit pairs.

\paragraph{Scale and shape.} With centering matrix
$P=I-\frac1k\mathbf{1}_k\mathbf{1}_k^\top$, $\|Ps\|/\sqrt{k}$ is the standard
deviation of the student's scores in the row and
$\rho=\langle Ps,Pt\rangle/(\|Ps\|\,\|Pt\|)$ their Pearson correlation with
the teacher's. Since $Pr=Ps-Pt$, expanding Eq.~\eqref{eq:center-loss} splits
PairMSE into a scale and a shape term:
\begin{equation}
\mathcal{L}_{\mathrm{center}}
=\tfrac{1}{k}\big(\|Ps\|-\|Pt\|\big)^2+\tfrac{2}{k}\|Ps\|\,\|Pt\|\,(1-\rho).
\label{eq:split}
\end{equation}
The first term matches the row's \emph{scale}, the second its \emph{shape}. When the teacher's scores in a row are nearly tied, $\|Pt\|$
is small, and the loss mostly asks the student to compress its score range. The per-row
Pearson loss keeps only the shape,
\(\mathcal{L}_{\rho}=\frac1k\big\|Ps/\|Ps\|-Pt/\|Pt\|\big\|^2=\frac2k(1-\rho)\),
and is unchanged by any shift or positive rescaling of either row. To separate
the two terms, we also train with a row-weighted Pearson loss
\(\mathcal{L}_{w}=\frac2k\,\mathrm{sg}(w)\,(1-\rho)\), where
$\mathrm{sg}$ stops the gradient through the weight $w$. Setting $w\equiv1$ gives
$\mathcal{L}_\rho$, while $w=\|Ps\|\,\|Pt\|$ gives the shape term of
Eq.~\eqref{eq:split} without its scale term.

\section{Experimental Setup}
\label{sec:setup}

\paragraph{Data.}
We use eight BEIR datasets~\citep{thakur2021beir}. Training uses 8,899
rows, one per query from the training splits of SciFact~\citep{wadden2020fact} (809),
NFCorpus~\citep{boteva2016fulltext} (2,590), and FiQA~\citep{maia2018fiqa}
(5,500). Each \emph{TF-IDF} row pairs one labeled relevant document with the
31 highest-ranked TF-IDF documents not labeled relevant (scikit-learn, English
stop words, 1--2-grams, 200,000 features, fitted on the collection's corpus).
Relevance labels are part of BEIR (qrels) and can mark several documents
relevant for one query.
\emph{Teacher-mined} rows instead take the 31 documents the teacher scores
highest after removing every document labeled relevant for the query and,
to drop likely unlabeled positives, every document scored above 95\% of the
positive's score \citep{moreira2024nvretriever}. The teacher's top document
is then the labeled positive in 99.9\% of rows; the rest are NFCorpus rows
with too few documents below the cut.
\emph{Label-free teacher} rows hold the teacher's 32 highest-scored documents
of the full corpus. For 67\% of queries the row contains the labeled positive (for 26\% it is the top document), and for 89\% it contains at least one document labeled relevant, the positive or another. Target-collection evaluation uses 1,271 test
queries from these three datasets, none shared with training. Transfer
evaluation uses 12,505 test queries from five excluded collections:
ArguAna~\citep{wachsmuth2018arguana}, SciDocs~\citep{cohan2020specter},
TREC-COVID~\citep{voorhees2020treccovid},
Webis-Touch\'e~2020~\citep{bondarenko2020touche}, and
Quora~\citep{thakur2021beir}. For every query we search the full corpus by
cosine similarity, exclude self-matches, and report
\ndcg~\citep{jarvelin2002cumulated} with equal weight per dataset.

\paragraph{Training.}
Both students train for one epoch with query/value LoRA ($r=64$,
$\alpha=128$) and AdamW (3\% warmup, cosine decay), at learning rates
$10^{-5}$ (Qwen) and $2\times10^{-5}$ (E5) and effective batches of 16 and 8
rows (557 and 1,113 steps), with inputs truncated to 512 tokens due to E5
limits;\footnote{Scoring documents at full length barely changes the
teacher's targets (median per-row Spearman 0.998); re-mining and retraining
with full-length scores moves no arm of Table~\ref{tab:loss}b by more than
0.001 on the target collections.} we report the
final checkpoint. Qwen models prefix queries with the instruction ``Given a web
search query, retrieve relevant passages that answer the query'', NV-Embed-v2
uses one query instruction per dataset, and E5 uses its ``query:'' and
``passage:'' prefixes; other documents are unprefixed. Merged LoRA
weights keep each student's architecture and cost. The label-only baseline trains a DPR-style
contrastive loss \citep{karpukhin2020dpr} over the row plus in-batch negatives
(256 candidates per softmax), with temperature and learning rate tuned on
NFCorpus and FiQA dev queries, on TF-IDF or teacher-mined rows. Pearson runs
change only the loss.

\paragraph{Teachers and statistics.}
The teacher is Qwen3-Embedding-8B unless stated otherwise;
Table~\ref{tab:main} adds an equal fusion of NV-Embed-v2 and
Qwen3-Embedding-8B. Gains are over the untrained student. Unless stated otherwise,
intervals are 95\% $t$ intervals over three seeds, holding test queries fixed.
Equivalence is tested with two one-sided tests (TOST) at margins $\pm0.005$ for TF-IDF controls and
$\pm0.01$ for the one-hot control. Per-dataset tests are paired $t$-tests with Holm correction.

\section{Results}

\subsection{Distillation Improves Compact Students}
\label{sec:main}

With Qwen3-Embedding-8B as the only teacher, distillation on TF-IDF rows
improves both students on the target collections by 0.018 \ndcg
(Table~\ref{tab:main}), closing 19\% and 18\% of the gap to the teacher,
significantly on all three datasets for Qwen and on NFCorpus and FiQA for E5.
Fusing in NV-Embed-v2 raises the target gains to 0.034 (Qwen) and 0.026 (E5),
but NV-Embed-v2 was fine-tuned on the target collections' training splits and
ArguAna \citep{lee2024nvembed}, and the Qwen3 Embedding report lists its
fine-tuning data only in part \citep{qwen2025embedding}. On the excluded
collections E5 gains 0.064, which suggests general transfer
(Section~\ref{sec:rows}); Qwen gains 0.012, 60\% of it from Touch\'e (49
queries).

\begin{table}[!t]
\caption{Macro \ndcg on target (3), excluded (5), and all eight collections;
TF-IDF rows, three-seed means, 95\% seed intervals for all-8. Teachers:
Qwen3-Embedding-8B, alone or fused with NV-Embed-v2$^\dagger$ (trained on the
target splits).}
\label{tab:main}
\centering
\small
\setlength{\tabcolsep}{3.5pt}
\begin{tabular}{@{}llrrr@{}}
\toprule
Student & Training & Target & Excluded & All-8 [seed 95\% CI] \\
\midrule
Qwen3 0.6B & none (base) & 0.506 & 0.575 & 0.549 \\
 & label-only & 0.505 & 0.534 & 0.523 [0.520,0.526] \\
 & distill., Qwen3-8B & 0.524 & 0.587 & 0.563 [0.562,0.565] \\
 & \hspace{0.5em}+ NV-Embed-v2$^\dagger$ & 0.540 & 0.586 & 0.569 [0.567,0.570] \\
\midrule
E5-large & none (base) & 0.500 & 0.480 & 0.487 \\
 & label-only & 0.501 & 0.483 & 0.490 [0.481,0.498] \\
 & distill., Qwen3-8B & 0.518 & 0.544 & 0.534 [0.529,0.539] \\
 & \hspace{0.5em}+ NV-Embed-v2$^\dagger$ & 0.526 & 0.545 & 0.538 [0.536,0.539] \\
\midrule
Qwen3 8B & teacher & 0.602 & 0.631 & 0.620 \\
\bottomrule
\end{tabular}
\end{table}

Tuned label-only training on the TF-IDF rows leaves the target collections
unchanged for both students, and Qwen loses 0.041 on the excluded collections; distillation with
either teacher beats it on all three averages. The merged students
encode queries 3.3$\times$ (Qwen) and 9.2$\times$ (E5) faster than the teacher
(evaluated on AMD MI210, batch 8).

\subsection{Teacher-Mined Negatives Need a Scale-Free Loss}
\label{sec:loss}

\begin{table}[!t]
\caption{Gain in \ndcg over the untrained student by row type and loss
(Qwen3-8B teacher; three seeds; --: not run). $\Delta$: difference from the
panel's reference row (ref.) on the target collections, paired by seed, with
its 95\% seed interval; Excl.: excluded collections. Other losses change only
the loss of the PairMSE runs; the one-hot target replaces the teacher's scores
by the label.}
\label{tab:loss}
\centering
\footnotesize
\setlength{\tabcolsep}{2.6pt}
\begin{tabular}{@{}lrrrrrr@{}}
\toprule
 & \multicolumn{3}{c}{Qwen3 0.6B} & \multicolumn{3}{c}{E5-large} \\
\cmidrule(lr){2-4}\cmidrule(lr){5-7}
Rows and training & Target & $\Delta$ & Excl. & Target & $\Delta$ & Excl. \\
\midrule
\multicolumn{7}{@{}l}{\emph{(a) TF-IDF rows, labeled}} \\
PairMSE (ref.) & $\mathbf{+.018}$ &  & $\mathbf{+.012}$ & $\mathbf{+.018}$ &  & $\mathbf{+.064}$ \\
Pearson & $+.017$ & $\phantom{+}.000{\pm}.001$ & $+.010$ & $+.023$ & $+.005{\pm}.003$ & $+.059$ \\
\midrule
\multicolumn{7}{@{}l}{\emph{(b) Teacher-mined rows, labeled}} \\
Pearson (ref.) & $\mathbf{+.030}$ &  & $\mathbf{+.012}$ & $\mathbf{+.037}$ &  & $\mathbf{+.059}$ \\
PairMSE & $+.006$ & $-.024{\pm}.006$ & $+.007$ & $+.023$ & $-.014{\pm}.001$ & $+.056$ \\
PairMSE, learned scale & $+.017$ & $-.013{\pm}.004$ & -- & $+.023$ & $-.014{\pm}.005$ & -- \\
$\mathcal{L}_w$, $w=\|Ps\|\,\|Pt\|$ & $+.024$ & $-.006{\pm}.003$ & -- & $+.033$ & $-.004{\pm}.002$ & -- \\
MarginMSE & $+.017$ & $-.013{\pm}.006$ & -- & -- &  & -- \\
KL, tuned temp. & $+.023$ & $-.007{\pm}.003$ & -- & -- &  & -- \\
Pearson, one-hot target & $+.034$ & $+.004{\pm}.004$ & $+.011$ & $+.040$ & $+.003{\pm}.002$ & $+.049$ \\
Label-only & $+.033$ & $+.003{\pm}.004$ & $+.018$ & $+.023$ & $-.014{\pm}.006$ & $+.056$ \\
\midrule
\multicolumn{7}{@{}l}{\emph{(c) Label-free teacher rows (teacher's top 32)}} \\
Pearson (ref.) & $\mathbf{+.025}$ &  & $\mathbf{+.014}$ & $\mathbf{+.033}$ &  & $\mathbf{+.064}$ \\
PairMSE & $+.018$ & $-.007{\pm}.002$ & $+.009$ & $+.032$ & $-.002{\pm}.004$ & $+.065$ \\
One-hot, teacher's top-1 & $+.016$ & $-.009{\pm}.003$ & $+.012$ & $+.027$ & $-.007{\pm}.006$ & $+.061$ \\
\bottomrule
\end{tabular}
\end{table}

Mining the teacher's hard negatives is the standard way to build such rows,
yet with PairMSE they give Qwen only a third of its TF-IDF gain, while
label-only training on the same rows gains more than five times as much
(Table~\ref{tab:loss}b). On mined rows the teacher's scores are nearly tied (median
within-row variance 0.0003, against 0.0079 on TF-IDF rows), so by
Eq.~\eqref{eq:split} PairMSE mostly asks for a compressed score range. On mined rows, PairMSE
shrinks Qwen's whole score scale to $\times0.34$ of the untrained student's,
while its standardized margin $z$ (the positive's score minus the mean
negative, divided by the negatives' standard deviation) barely rises.

Changing only the loss to per-row Pearson raises the target gain on
the same rows, to $+0.030$ for Qwen and $+0.037$ for E5
(Table~\ref{tab:loss}b). Qwen keeps its scale, and its
$z$ on mined rows rises to 2.11, $+0.21$ [$+0.09$, $+0.33$] above PairMSE's
(interval over rows), level with label-only training (2.09). With this change Qwen matches
label-only training on these rows, although Pearson sees 32 candidates per
query against the baseline's 256, and E5 exceeds it.
On the excluded collections Pearson trails label-only training
for Qwen ($+0.012$ vs.\ $+0.018$). Pearson also beats standard alternatives:
listwise KL \citep{tamber2025listwise} with student temperature 0.05 and a
teacher temperature tuned on dev queries (0.025 of $\{0.0125, 0.025, 0.05\}$)
and MarginMSE \citep{hofstatter2020marginmse} give Qwen $+0.023$ and $+0.017$,
below Pearson by $0.007$ and $0.013$ over three seeds (Table~\ref{tab:loss}b);
the KL gap is mostly on NFCorpus. Pearson's gradients start about $8\times$ larger, but gradient
clipping never triggered; we did not retune PairMSE's learning rate.

Further controls isolate the scale term (target collections). Multiplying the student's scores by one
trainable scalar before PairMSE (shared by all rows, dropped at
retrieval) raises Qwen's gain from $+0.006$ to $+0.017$,
about half of Pearson's lead; the
scalar settles at 0.36, close to the $\times0.34$ by which plain PairMSE
shrinks Qwen's scores. For E5 the scalar stays near 1 and changes nothing.
Removing the scale term in every row instead, by training with
$\mathcal{L}_w$, $w=\|Ps\|\,\|Pt\|$ (Section~\ref{sec:method}; $w$ divided
by a fixed constant), recovers 76\% [66\%, 86\%] of Pearson's lead
for Qwen and 75\% [61\%, 90\%] for E5 (Table~\ref{tab:loss}b); $w=\|Pt\|$
recovers 90\% and 97\% (one seed). So the scale term, not the row weighting, accounts for most of the gap:
globally for Qwen and row by row for E5, whose untrained rows match the
teacher's scale on average, so one scalar has no average shrink to
correct. Both students keep a small residual ($-0.006$ and $-0.004$), for
Qwen largest on NFCorpus, whose teacher rows are the flattest, consistent with
Pearson's equal weight per row also mattering there.

\begin{figure}[t]
\centering
\includegraphics[width=\columnwidth]{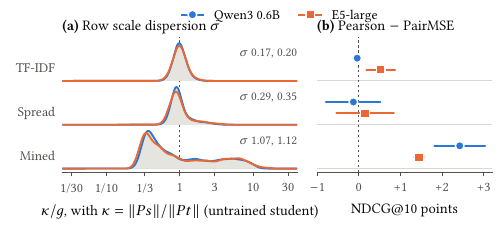}
\caption{Row scale dispersion separates the rows where the Pearson loss works better.
(a) Per-row $\kappa=\|Ps\|/\|Pt\|$ of the untrained students over its geometric
mean (900 training rows per type); $\sigma$ is its log-space standard
deviation. (b) Pearson $-$ PairMSE on the target collections, 95\% seed intervals
(spread rows: one seed, query-bootstrap interval).}
\label{fig:dispersion}
\end{figure}

This suggests a diagnostic computed before training, \emph{row scale
dispersion} $\sigma$ (Figure~\ref{fig:dispersion}): the standard deviation of $\log \kappa$ over rows, with
$\kappa=\|Ps\|/\|Pt\|$ for the untrained student. One learned scalar can absorb the
geometric mean $g$ of $\kappa$, but only a per-row scale-free loss removes $\sigma$.
On mined rows, $1/g$ is 0.30 for Qwen, near the learned 0.36, and $g=0.99$ for
E5, whose scalar did nothing. $\sigma$ is about 1.1 on mined rows and 0.2 on
TF-IDF rows, where the losses tie for Qwen and differ by only $0.005$ for E5
(Table~\ref{tab:loss}a). On mined rows the two students' per-row $\log\kappa$
correlate at 0.98, so on these rows $\sigma$ reflects the rows more than the
student. Rows sampled across the teacher's ranking, the remedy
of \citet{jang2026beyond}, sit in between ($\sigma\approx0.3$; PairMSE gains
$+0.013$ and $+0.009$), and the gap between the losses is small:
$-0.001$ (Qwen) and $+0.002$ (E5; one seed).

\subsection{On Mined Rows, Teacher Scores Replace Labels}
\label{sec:teacher}

Because mining starts from the labeled positive, the teacher's scores on mined
rows largely restate the label: on sampled training rows, the teacher's row
correlates with the one-hot label vector at median 0.74 (54\% of its variance),
against 0.35 (14\%) on TF-IDF rows. Replacing them with a one-hot target on the
positive, under the same Pearson loss, rows, seeds, and code, does at least as
well (Table~\ref{tab:loss}b): the two are equivalent within
$\pm0.01$, and within $\pm0.005$ for E5. The one-hot target also beats label-only
training for E5, by $0.017$. On these rows the teacher's
contribution is thus the choice of negatives, and the gain comes from a
scale-free margin objective on the label; the teacher's scores help only E5's
excluded collections ($+0.009$ [$+0.007$, $+0.012$]). On TF-IDF rows, which
the teacher did not select, the same one-hot target instead harms both
students ($-0.043$ and $-0.048$, one seed), while the teacher's scores help
($+0.017$ and $+0.023$): a one-hot target asks for tied negatives, which hurts
on these rows, as flat targets do (Section~\ref{sec:ordering}).

On unlabeled datasets, the teacher can provide surrogate labels. On label-free
teacher rows (Table~\ref{tab:loss}c), Pearson gains $+0.025$ (Qwen) and
$+0.033$ (E5) over three seeds: 82\% and 90\% of
its gain on labeled mined rows. Much of this gain needs only the teacher's
top document: a one-hot pseudo-label on it (labeled relevant in 54\% of rows)
gains $+0.016$ (Qwen) and $+0.027$ (E5) under the same loss, 65\% and 79\%
of the scores' gain; the full scores add a clear gain for both students, and are equivalent on the
excluded collections. Unlike the label on TF-IDF rows,
a one-hot target on rows the teacher chose does not collapse. The teacher's own top 32 are less tied than mined rows
(median variance 0.0016 vs.\ 0.0003), so the loss matters less here: Pearson
beats PairMSE by only $0.007$ for Qwen and not measurably for E5.

\subsection{Target-Collection and Label Controls}
\label{sec:rows}

Two controls check the TF-IDF result of Section~\ref{sec:main}. Rows built
from MS MARCO \citep{nguyen2016msmarco} instead of the target collections give
no target gain (one seed: $+0.000$ Qwen, $-0.001$ E5), so we need to use  
target in-domain rows for best results; E5 keeps most of its excluded gain (0.047 vs.\ 0.061),
although this control changes the domain as well as the collection. Since
PairMSE never uses the label, we also built rows without it (a query's top 32
TF-IDF candidates): they change Qwen's target score by only $-0.002$,
equivalent within $\pm0.005$ over three seeds (E5, one seed:
$-0.001$).

\subsection{Which Parts of the Score Vector Matter}
\label{sec:ordering}

Does the student learn from the positive's margin or from the negatives' own
scores? On TF-IDF rows we also train on the margin alone (loss $m^2$, with $m$
the student's error on the teacher's positive-minus-mean-negative margin). Over three seeds the margin is
enough for Qwen on the target collections, keeping 95\% of its full-vector
gain (equivalent within $\pm0.005$), but not for E5, which keeps none (0.020
below the full vector). On the excluded collections, dropping the negatives' scores costs both
students part of their gain (Qwen $-0.008$, E5 $-0.022$, both significant), so
those scores carry much of the transfer. Corrupted scores hurt more than missing
ones: shuffling the negatives' scores or setting them to their mean drops Qwen
0.05--0.07 below the untrained student (one seed).

\section{Conclusion}

How the training rows are built determines what an embedding teacher's scores
can teach. When the teacher mines negatives around a labeled document (here,
its top 31 below 95\% of the positive's score), its scores on them are nearly
tied and PairMSE mostly teaches the student to shrink its scores, which row
scale dispersion flags before training. A per-row
Pearson loss avoids this, bringing Qwen level with label-only contrastive
training on the same rows and lifting E5 above it. On these rows the scores add
little to the label: the teacher's value lies in choosing the negatives and,
without labels, mostly in its top document. The practical rule: use a
scale-free loss where row scale dispersion is high, as on teacher-mined
negatives, and on such rows use teacher scores only where labels are missing.

\paragraph{Limitations.}
We use one teacher, three target collections, one mining rule, and 32-document
rows; corrupted targets, the $w=\|Pt\|$ variant, the spread-row comparison
and the scale measurements use one seed.
The fused teacher gain may be smaller on unseen collections,
label-free rows still use benchmark queries, and we do not compare with a
distilled query encoder over teacher document embeddings
\citep{kim2023embeddistill,leaf2025teacheraligned}.

\section*{Ethical Considerations}

We use public datasets and teacher-generated scores; label-free rows from
real query logs would need privacy review. Students can inherit a teacher's
ranking biases, so deployments should check for systematic relevance errors;
NV-Embed-v2 is licensed CC BY-NC 4.0.

\ifdefined\ARXIV
\begin{acks}
We thank Supriyo Ghosh and Writabrata Bhattacharya for their questions about
embedding-model fusion. Those questions motivated the initial experiments
that grew into this work.
\end{acks}
\fi

\clearpage
\balance
\bibliographystyle{ACM-Reference-Format}
\bibliography{references}

\end{document}